\documentclass[twocolumn,aps,prl,groupaddress,superscriptaddress,showpacs]{revtex4-1}
\usepackage[T1]{fontenc}
\usepackage[latin9]{inputenc}
\usepackage{color}
\usepackage{graphicx}
\usepackage{amssymb}
\usepackage{multirow}

\makeatletter

\@ifundefined{textcolor}{}
{%
 \definecolor{BLACK}{gray}{0}
 \definecolor{WHITE}{gray}{1}
 \definecolor{RED}{rgb}{1,0,0}
 \definecolor{GREEN}{rgb}{0,1,0}
 \definecolor{BLUE}{rgb}{0,0,1}
 \definecolor{CYAN}{cmyk}{1,0,0,0}
 \definecolor{MAGENTA}{cmyk}{0,1,0,0}
 \definecolor{YELLOW}{cmyk}{0,0,1,0}
 }

\makeatother

\usepackage{amsmath}
\usepackage{amsfonts}%
\usepackage{graphicx}
\usepackage{dcolumn}
\usepackage{bm}
\usepackage[mathscr]{eucal}
\usepackage{array}

\newcommand{\Tr}{\mbox{tr}}

\newcommand{\req}[1]{Eq.~(\ref{#1})}
\newcommand{\reqs}[1]{Eqs.~(\ref{#1})}
\newcommand{\rref}[1]{(\ref{#1})}

\renewcommand{\r}{\mathbf{r}}

\renewcommand{\k}{\mathbf{k}}

\newcommand{\htau}{\hat{\tau}}
\newcommand{\beq}{\begin{equation}}
\newcommand{\eeq}{\end{equation}}
\newcommand{\be}{\begin{equation}}
\newcommand{\ee}{\end{equation}}
\newcommand{\beqa}{\begin{eqnarray}}
\newcommand{\eeqa}{\end{eqnarray}}
\newcommand{\bea}{\begin{eqnarray}}
\newcommand{\eea}{\end{eqnarray}}

\newcommand{\opsi}{{\Psi}}
\newcommand{\opsid}{{\Psi}^\dagger}

\newcommand{\hh}{\hat{h}}
\newcommand{\hM}{\hat{M}}

\begin{document}


\title{Spontaneous symmetry breaking and Lifshitz transition in
  bilayer graphene}

\author{Y. Lemonik}
\affiliation{Physics Department, Columbia University, New York, NY
  10027, USA }


\author{I.L. Aleiner}
\affiliation{Physics Department, Columbia University, New York, NY
  10027, USA }
\affiliation{Kavli Institute for Theoretical Physics China, CAS, Beijing 100190, China}

\author{C. Toke}
\affiliation{Physics Department, Lancaster University, Lancaster, LA1 4YB, UK}

\author{V.I. Fal'ko}

\affiliation{Kavli Institute for Theoretical Physics China, CAS, Beijing 100190, China}
\affiliation{Physics Department, Lancaster University, Lancaster, LA1 4YB, UK}

\begin{abstract}
We derive the renormalization group equations describing all the
short-range interactions in bilayer graphene allowed by symmetry and
the long
range Coulomb interaction. For certain range of parameters, we predict
the first order phase transition to the uniaxially deformed gapless state
accompanied by the change of the topology of the electron spectrum.
\end{abstract}
\pacs{73.22.Pr, 73.21.-b}
\maketitle

{\em Introduction}-- The Lifshitz transition (LiTr) \cite{Lifshitz} is the
simplest topological effect in
physics of metals. It consists of the change
of connectivity of isoenergetic surfaces, either as a function of
electron density or external parameters, such as strain. 
As the change of the topology of the {\em e.g.} Fermi surface can not
be continuous, all the observables in the system should experience
singularities at the LiTr also known as a half-integer-order
phase transition (PT). Alternatively,
the reconstruction of the Fermi surface may occur via an underlaying
spontaneous symmetry breaking PT. 
The observation of the LiTr in the bulk metals is an
extremely challenging task: a variation of the Fermi level in metals
requires doping which introduces disorder and obscures the transition,
whereas application of strain requires  high pressure experiments.

\begin{figure}[h]
\includegraphics[width=\columnwidth]{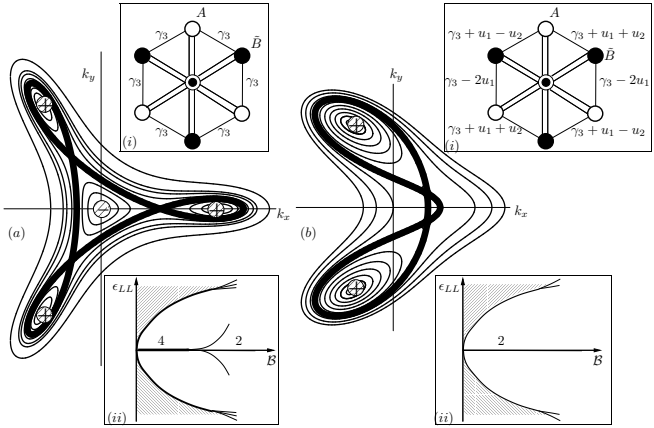}
\caption{
\label{fig1}
Constant energy lines for the one particle spectrum in the BLG. LiTr as a function of density occurs when the
Fermi level intersects the separatrix (bold line) for
a) unbroken ${\cal C}_{6v}$ symmetry of the graphene lattice;
b) symmetry breaking   ${\cal C}_{6v} \to {\cal C}_{2v}$. Circles mark Dirac points and $\pm$
indicates Berry phases, $\phi_B=\pm \pi$. Insets shows the tight-binding
cartoons for  the band structure (i) and the schematic
evolutions of the Landau levels in the magnetic field (ii) (the number
indicates the degeneracy per one spin and valley). 
}
\end{figure}

Bilayer graphene (BLG) -- a two-dimensional allotrope of carbon with a
honeycomb lattice -- is  a potentially ideal system to study  the
LiTr \cite{Falko}.
{ 
A gapless low-energy electronic structure of the conduction and valence
bands near the Brillouin zone (Bz) corners in Bernal stacked
BLG
has a parabolic dispersion $\epsilon_\pm\approx\pm p^2/2m$ at intermediate energies
determined by the  intra- and interlayer hops between closest
neighbors. Remarkably, the  electronic wavefunctions
accumulate the $\phi_B=2\pi$ Berry phase as the momentum going along
the loop encompassing $p=0$. This 
causes the double degeneracy (per one spin and one
valley) of the zero-energy Landau level (LL) in the magnetic field.
Those features, however, are not protected by the crystal symmetry. 
The parabolic dispersion is trigonally deformed at the lowest energies and
$\epsilon=0$ state splits into four Dirac points: one in the Bz corner
and three off-sets separated by
momentum $2mv_3$  due to the next-neighbor interlayer hopping, 
Fig.~\ref{fig1}a (this separation is about $0.1\%$ of the size of Bz). The total $\phi_B$
is conserved so that the Dirac point in the Bz corner carries  $\phi_B=-\pi$ and each of the off-sets $\phi_B=\pi$. This doubles
the degeneracy of zero-energy LL at weak magnetic
field.}  

Unlike in conventional metals, the LiTr in the BLG
 (from one to four Fermi lines) can be tuned by a small density variation controlled by a gate
voltage. Suitable suspended BLG devices of
sufficient quality for the LiTr studies  have been 
fabricated \cite{Yacoby,Geim}.
 
The central question of this Letter in the stability of the above
Lifshitz transition against the effect of  electron-electron
interaction (EEI). 
There are two possibilities: (i) EEI does not break the symmetry
leading to a quantitative renormalization of the band structure
affecting, {\em e.g.} the density {$n_{LiTr}=(2/\pi^2)(mv_3/\hbar)^2$} corresponding to
the LiTr;
(ii) EEI does break the symmetry leading to a qualitative
transformation of the spectrum -- the number of the Dirac points is
then determined by the reduced symmetry  (contradicting scenarios were
suggested 
in Refs.~\cite{Vafek,Levitov1,Levitov2} ).
Using the renormalization group (RG) treatment of the
EEI problem we found that: (i) $n_{LiTr}$ is not renormalized; (ii)  
the most likely spontaneous symmetry breaking in BLG 
occurs by the generation of the asymmetric hopping in the effective Hamiltonian
with the same symmetry as the effect of $A-\tilde{B}$ sublattice displacement
~\cite{Vafek}, see
Fig.~\ref{fig1}b. 
For the gedanken experiment where  $v_3$  is
varied, the symmetry breaking occurs via first order quantum PT after which the spectrum remains gapless but two Dirac
points are annihilated and two other persist and carry $\phi_B=\pi$;
 as the result, the degeneracy of the zero-energy Landau level is
half that of the  for unbroken ${\cal C}_{6v}$ symmetry.
The Berry phases control
the degeneracy of the zero-energy LL which is visible via
the Shubnikov-de Haas oscillations, and for some range of parameters
the finite temperature PT is of the first
order leading to bistabilities in transport.


{\em The low energy model} 
for the bilayer graphene is formulated in terms of the states
close to $K$ and $K'$ points of the Brillouin zone \cite{Falko}.
The Hamiltonian is
\begin{subequations}
\label{h-eff}
\be
\hat{\cal H}\!=\!\int d^2\r\opsid_\sigma\left[\hh_0+
\hh_w +\hh_c +\hh_{sr} \right]\opsi_\sigma.
\label{h}
\ee
Hereinafter, the summation over repeated spin indices $\sigma=\pm
1/2$ is implied.  Four component fermionic field $\opsi_\sigma=\left(\hat{\psi}^{A,K}_\sigma,
\hat{\psi}^{\tilde{B},K}_\sigma; \hat{\psi}^{\tilde{B},K'}_\sigma, -\hat{\psi}^{A,K'}_\sigma
 \right)$ 
lives in the valley ($KK'$) and the sublattice ($A\tilde{B}$) spaces \cite{Falko}
(sublattices $A$ and $\tilde{B}$ belong to the different layers).
All matrices acting in this four dimensional space are represented
as direct product of the Pauli matrices $\htau_i^{A\tilde{B}},\htau_i^{KK'}$,
(i=0,1,2,3):
\be
\hM_i^j\equiv\htau_i^{KK'}\otimes\htau_j^{A\tilde{B}}
\ee
and $\htau_0^{\dots}$ is the unit $2\times 2$ matrix.

The kinetic energy is given by ($\hbar=1$, $k_{x,y}=-i\partial_{x,y}$)
\be
\hh_0(k_{x,y})=
\left[\hM_3^1\left(k_x^2-k_y^2\right)
 \!\! - 2\hM_3^2k_x k_y \right]/(2m).
\label{kin}
\ee
Together with \req{kin}, the trigonal warping term,
\be
\hh_w(k_{x,y})=-v_3
\left[\hM_0^1k_x+\hM_0^2k_y \right],
\label{kin-w}
\ee
 determines the spectrum in Fig.~\ref{fig1}a.

 The
long-range Coulomb interaction, 
\be
\hh_{c}={{e^2} \over 2}
\int
{{ d^{2}\r^\prime\opsid_{\sigma^\prime}(\r^\prime)\opsi
_{\sigma^\prime}(\r^\prime)} \over {|\r-\r^\prime|}},
\label{Coulomb}
\ee
is the strongest in the system. 
However, due to the screening it does not scale and therefore does not describe any symmetry
breaking by itself.
The latter is captured by the scaling of the marginal short-range interaction
\be
\hh_{sr}=
({2\pi}/{m})\sum_{i,j=0}^3 g_i^j \hM_i^j
\left[\opsid\hM_i^j \opsi\right].
\label{shortrange}
\ee
The couplings $g_{i}^{j}$ are not independent \cite{AKT}. The ${\cal
C}_{6v}$ symmetry of the bilayer constrains
\be
\begin{split}
&g_{1}^{1}=g_{2}^{2}=g_{1}^{2}=g_{2}^{1}=g_{G};\quad
g_{3}^{1}=g_{3}^{2}=g_{E_1};\\
&g_{1}^{3}=g_{2}^{3}=g_{E_1^{\prime\prime}}; \quad g_{0}^{1}=g_{0}^{2}=g_{E_2};
\quad g_{1}^{0}=g_{2}^{0}=g_{E_2^{\prime\prime}};\\
&g_{0}^{3}=g_{B_1};\quad g_{3}^{0}=g_{A_2};\quad g_{3}^{3}=g_{B_2},
\label{symmetries}
\end{split}
\raisetag{-20pt}
\ee
\end{subequations}
where subscripts indicate the irreducible representations of the 
extended point group, see {\em e.g.} Sec. III of Ref.~\cite{Basko}.
For example, $E_1$ is two dimensional representation which does not
change sign under $C_2$ rotation and describes the symmetry breaking
shown on Fig.~\ref{fig1}b, whereas $B_2$ is the one dimensional
representation describing the breaking of the interlayer symmetry,
${\cal C}_{6v} \to {\cal C}_{3v} $.

\begin{figure}[h]
\includegraphics[width=1\columnwidth]{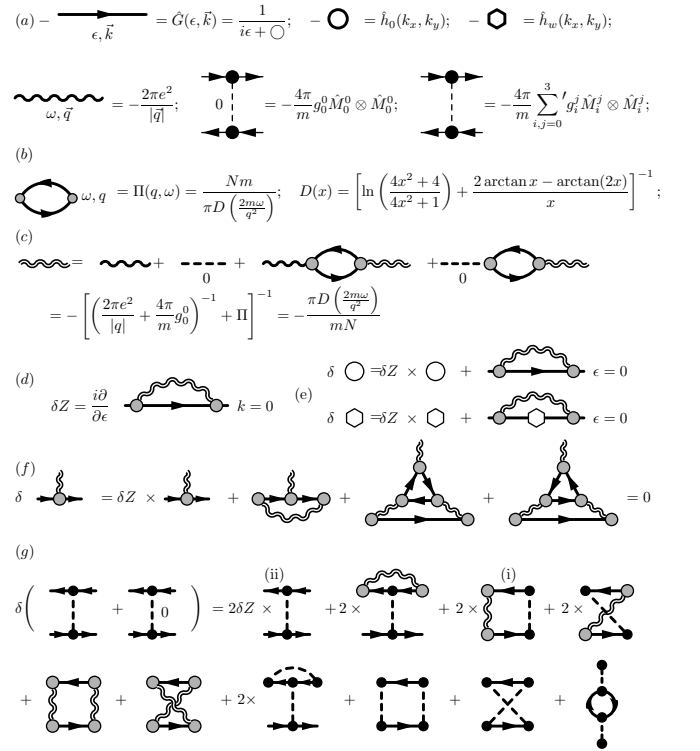}
\caption{Derivation of the RG equations. a) Definitions of the
  elements; b) Polarization operator; c) Screening of the Coulomb
  interaction;
d,e) Renormalization of one particle spectrum; f) gauge invariance of the
scalar vertex; g) renormalization of the short range interaction}
\label{fig2}
\end{figure}

{\em The RG study} of the model
\rref{h-eff} is based upon the analysis of diagrams shown in Fig.~\ref{fig2}.
The Coulomb interaction apparently is the most relevant
operator ({\em i.e.} its perturbative treatment leads to the linear rather
than the logarithmic divergence). The
screening of this interaction, see Fig.~\ref{fig2}c, makes it 
marginal; its  value is $\simeq 1/[N\Pi
(q,\omega)]$,
($\Pi$ is the polarization operator).
The formal justification for the approximation
Fig.~\ref{fig2}c is the $1/N$ expansion -- which we believe is applicable
for $N=4$ -- and the long wavelength limit.
Note, that $g_{0}^{0}$ enters together with the Coulomb
interaction 
potential so that it drops out, see Fig.~\ref{fig2}c, 
and does not contribute to the running of the
 coupling constants. 
Other constants $g_i^j$ are assumed to be small and treated in a first
loop approximation \cite{Son}.

\begin{subequations}
Because the polarization operator does not have logarithmic divergences and
all of the interactions are considered in the first loop, the details of the
cut-off scheme are not important. On each step we will restrict the internal
momentum of the loop as ${\cal E}-d{\cal E}< k^2/2m({\cal E})
\leq  {\cal E}$.
We, then, rescale $\psi \to \left(1+\delta Z/2\right)\psi$, ($\delta Z$ is defined
on Fig.~\ref{fig2}d) to keep the term $\partial_\tau\psi$ in the
Matsubara equation intact. As a bonus, the scalar vertex is also not
renormalized for the reason of gauge invariance, see Fig.~\ref{fig2}f.
\label{RG}
Renormalizations of $m$ and $v_3$, see Fig.~\ref{fig2}e, are given by\cite{Footnote}
\be
{d\ln m}/{d\ell}=-{d\ln v_3}/{d\ell}=-{\alpha_1}/{N}; \ \alpha_1\approx -.078, 
\label{mass}
\ee
where $\ell\equiv \log({\cal E}_0/{\cal E})$, where ${\cal E}_0\simeq 0.3\, eV$ limits the
applicability of the two-band model of bilayer graphene. 

The possible symmetry breakings are described by the scaling of the
short range interaction terms \cite{Footnote}:
\be
\begin{split}
&\frac{dg_{i}^{j}}{d\ell}=-\frac{\tilde{\alpha}\delta(E_1)_{i}^{j}}{N^2}-
\frac{\alpha_1 g_{i}^{j}}{N}-NB_{i}^{j}\left(g_{i}^{j}\right)^2-\!\!\!\!\!\!\!\sum_{k,l,m,n=0}^3\!\!\!\!\!\!
C_{i;km}^{j;ln}
\tilde{g}_{k}^{l}\tilde{g}_{m}^{n}
\\
&
\tilde{g}_{i}^{j}\equiv g_{i}^{j}(1-\delta_{i0}\delta_{j0}) +
\delta_{i0}\delta_{j0}{{\alpha_2}/{(2N)}},
\ \alpha_2\approx .469
\end{split}
\raisetag{14pt}
\label{RGgs}
\ee
where $\tilde{\alpha}=\alpha_3 - \alpha_2^2/16,\ \alpha_3\approx .066$, the symbol
$\delta(E_1)_{i}^{j}$ is 
defined as  $\delta(E_1)_{i=3}^{j=1,2}=1$ and  $\delta(E_1)_{i}^{j}=0$ otherwise. The summation over
repeated indices is not implied in \req{RGgs}.
The constants in \req{RGgs} are given
by
\be
\begin{split}
B_{i}^{j}&=
\frac{1}{16}\sum_{l=1,2}{\Tr}\left\{\left[\hat{M}_i^j,
\hat{M}_3^l\right]^2\right\};\\
C_{i;km}^{j;ln}&=
\frac{1}{32}\left\{\Tr
\left[\hat{M}_i^j\left[\hat{M}_k^l,\hat{M}_m^n\right]\right]
\right\}^2
\\
&+\frac{1}{64}\sum_{r=1,2}\left\{\Tr \hat{M}_3^r
\left(\hat{M}_k^l\hat{M}_i^j\hat{M}_m^n + \hat{M}_m^n\hat{M}_i^j\hat{M}_k^l\right)
\right\}^2
\\
&+\frac{\delta_{ik}\delta_{jl}}{4}\sum_{r=1,2}\Tr\left\{\hat{M}_m^n\hat{M}_3^r
\left[\hat{M}_i^j,\hat{M}_3^r\right]
\hat{M}_m^n  \hat{M}_i^j\right\}.
\end{split}
\raisetag{81pt}
\label{constants}
\ee
Note that,  \reqs{RGgs} and \rref{constants} respect
symmetry \rref{symmetries}.


Equations \rref{RG} are the main technical result of this paper. They
describe the evolution of all the band structure parameters and all short-range interactions terms
allowed by symmetry in the
leading logarithmic approximation. To compare with the existing
literature:
the RG treatment of Ref.~\cite{Vafek} considers only two possible terms ($g_{E_1}$, and
$g_{B_1}$ ), treats the Coulomb interaction as shortrange and
neglects the warping in the spectrum; mean-field treatment of
Refs.~\cite{Levitov1,Levitov2} corresponds to hardly justifiable taking into account only
one \cite{Levitov1} [(i) of Fig.~\ref{fig2}g] or two \cite{Levitov2} diagrams [(i,ii) of
Fig.~\ref{fig2}g] with the subsequent projection on the $B_2$
representation.

\end{subequations}

{\em  RG flow and non-broken symmetry} --
The density of electrons (or holes) at which the topology of the Fermi
{surface changes is found from \reqs{kin} and \rref{kin-w} as
$n_{LiTr}=({2}/{\pi}^2) \left({mv_3}/{\hbar}\right)^2
\simeq 2 \times 10^{10}cm^{-2}$ (estimated with $m=0.035$ and $v_3=
v_3\simeq 10^7cm/s$ ). }
According to \reqs{mass}, the Coulomb part of the EEI does not
renormalize $n_{LiTr}$ 
but affects the energy of the saddle points in the
single-particle spectrum 
${\cal E}_{LiTr}\equiv {mv_3^2}/{2}$.
The bare { value of this energy can be estimated using  the bilayer parameters
  $m,\ v_3$ quoted above
as ${\cal E}_{LiTr}\simeq 1\, {\rm meV}$.} The renormalized value is
$
\tilde{\cal E}_{LiTr}={\cal E}_{LiTr}\left({{\cal E}_0}/{{\cal
      E}_{LiTr}}\right)^{\frac{\alpha_1}{N}} \simeq {\cal E}_{LiTr}\left({{\cal
      E}_{LiTr}}/{{\cal E}_0}\right)^{0.02}, $
such change is not observable.

{\em RG flow and symmetry breaking} --
The divergence of a coupling constant $g_{\cdot}$ during the renormalization
signals
the symmetry breaking with the order parameter from the corresponding irreducible
representation (a more complete classification, involving the 
the magnetic and gauge symmetries will be reported elsewhere
\cite{elsewhere}). 

Let us assume with the short-range interactions on the energy scale
$E_0$ are negligible, $g_{i}^j=0$.
The constant term in \req{RGgs} means that this point
is not fixed
and  couplings $g_{E_1}$ and $g_{B_1}$ (see \req{symmetries}) will flow away
from this point.
Ignoring $g_{B_1}$, we obtain an equation for $g_{E_1}$:
\be
\begin{split}
&\frac{dg_{E_1}}{d\ell} = -\frac{c_1}{N(N+2)}-2(N+2)\left(g_{E_1} - c_2\right)^2;
\\
&c_1\approx
\left(\frac{\alpha_3(N+2)}{N} -\frac{(\alpha_2-\alpha_1)^2}{8N}\right);\ c_2\approx \frac{\alpha_2-\alpha_1}{4N(N+2)}.
\end{split}
\raisetag{1.4cm}
\label{glFlow}
\ee
Note that $c_1>0$, for $N>0$, and no 
fixed point exists;  though $\alpha_3$
appears small, its neglecting  would lead
to a  nontrivial fixed point $g_{E_1}\simeq 1/N^3$.
Solution of \req{glFlow} is
\be
g_{E_1}(\ell) = c_2 - \sqrt{\frac{c_1}{2N(N+2)^2}}
\cot\left[\sqrt{\frac{2c_1}{N}}(\ell_0-\ell)
\right]
\label{glSol}
\ee
where $\ell_0$ is found from  $g_{E_1}(\ell=0)=0$: for $N=4$
$\ell_0 \sim 7.1$.
Inclusion of $g_{B_1}$ shifts the pole slightly so that $(g_{E_1};\
g_{B_1})\propto ({\ell_0-\ell})^{-1}
(-1.67; 0.85)
$.

This divergence implies a symmetry breaking at ${\cal E}_{E_1}\simeq E_0e^{-7.1}
\simeq  0.3 {\rm meV}$ \cite{footnote2}.
It is important to notice that ${\cal E}_{E_1}$ and ${\cal E}_{LiTr}$ turn out to be
of the same order and, therefore, have to be considered together.
A more accurate theoretical comparison of those two energy scales
requires more detailed knowledge about the microscopic values of the
initial interaction constants which is not available at this time. 
Therefore, we will discuss the possible PTs for
an arbitrary value of $\Upsilon\equiv{\cal E}_{E_1}/{\cal
  E}_{LiTr}$. 

\begin{figure}[h]
\includegraphics[width=\columnwidth]{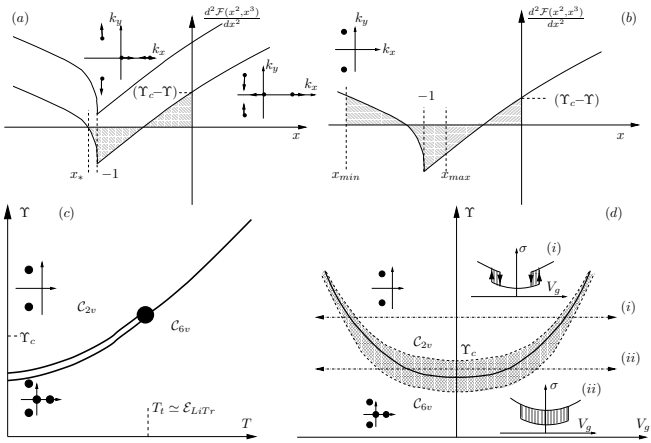}
\caption{a,b)The curvature of the mean-field energy along the steepest
  decent direction. The square root singularity is caused by the
  collision of the Dirac points shown on the inset. At $\Upsilon <
  \Upsilon_1 < \Upsilon_c$ extra local minimum and maximum are formed (a), 
 such as $|u_1^{max}|> {\cal E}_{LiTr}$ (b), i.e. only two Dirac points
  remain. The total dashed area equals to zero. c,d) The schematic phase
diagrams for the finite temperature (c) and for quantum (d) (controlled by
the gate voltage $V_g$)  PTs. The insets show predicted hysteretic (or
slow noise)
behavior (dashed areas) of the conductivity for the corresponding paths on the phase diagram. }
\label{fig3}
\end{figure}

{\em Analysis of the phase transition}--
If  $\Upsilon\lesssim 1$, the divergence of
$g_{E_1}$ is terminated and the symmetry is not broken.
The possible divergence of the coupling constant $g_{E_1}$ at
$\Upsilon \gtrsim 1$ indicates the symmetry breaking and
appearance of the anomalous averages comprising the irreducible
representation $E_1$ of the group ${\cal C}_{6v}$, 
\be
u_{j} = ({2\pi}/{m})\langle\opsid_\sigma
\hM^{j}_3\opsi_\sigma\rangle;\quad j=1,2.
\label{us}
\ee
For studying the PT we have to consider the Landau
free energy density. It
must be of the form
\be
f=n_{LiTr}{\cal E}_{LiTr}{\cal F}_{\Upsilon}\left(\frac{u_1^2+u_2^2}{{\cal
      E}_{LiTr}^2};
\frac{u_1^3-3u_1u_2^2}{{\cal
      E}_{LiTr}^3}\right)
\label{f}
\ee
for the symmetry and dimensionality reasons.
At $\Upsilon < \Upsilon_c$ (here $\Upsilon_c\simeq 1$), 
function ${\cal F}(x,0)$ has a local minimum at $x=0$
which, at $\Upsilon > \Upsilon_c$,  turns to a maximum.
The presence of the cubic invariant prescribed by ${\cal C}_{6v}$ symmetry
signals that the zero temperature PT,   
under varying $\Upsilon$, can be only of the first order and occurs at
$\Upsilon < \Upsilon_c$. 

Now, we argue that the value of the order parameter in the ordered
phase is  such that the electron spectrum has two Dirac points, as in
Fig.~\ref{fig1}b. In the mean-field approximation, the one particle
Hamiltonian reads
\be
\hat{H}=\hh_0(k_{x,y})+\hh_w(k_{x,y})-u_1\hM_3^1 - u_2\hM_3^2.
\label{MFa}
\ee
At $u_1=-{\cal E}_{LiTr}, u_2=0$  two Dirac points collide and
disappear and the band structure of Fig.~\ref{fig1}b is formed. At 
$u_1=3{\cal E}_{LiTr}, u_2=0$  three Dirac points collide, and, once
again, the spectrum with two Dirac points is formed.

The mean-field energy density is given by
\be
\begin{split}
&f_{MF}(u_1,u_2)=\left(\Upsilon_c-\Upsilon\right)\frac{m u_iu_i}{2\pi}
\\
&\
+\int\frac{d^2k}{\pi^2}
\left[
\epsilon(\k)- \epsilon(\k)\vert_{u_{1,2}=0}
-\left.
\frac{u_iu_j}{2}\frac{\partial^2\epsilon(\k)}{\partial u_i\partial u_j}
\right\vert_{u_{1,2}=0}
\right],
\end{split}
\raisetag{1.7cm}
\label{MFB}
\ee
where $\epsilon$ is the negative eigenvalue of $\hat{H}$, see \req{MFa},
and the summation over the repeated indices $i,j=1,2$ is implied.
The curvature of the energy density, found from \req{MFB}, see
 Fig.~\ref{fig3},  indicates that indeed $u_1$
formed during the PT transforms spectrum of Fig.~\ref{fig1}a to that
of Fig.~\ref{fig1}b.

The corrections to the mean-field can not remove the singularity for
the colliding Dirac points, as the Hamiltonian \rref{MFa} at low
energies is protected by symmetry. Therefore, we believe, that our
conclusion about the number of Dirac points in ordered and disordered
phases is more general than the mean-field derivation.

The conclusion about the first-order quantum phase transition have
the important consequences for the finite temperature phase
diagram, see  Fig.~\ref{fig3}c. At low temperatures $T \ll
{\cal E}_{LiTr}$, the transition remains of the first order up to some
tricritical temperature $T_t$, and at $T_t$ the transition is continuous
and belongs to $3$ states Potts model universality class \cite{Vafek}.
The quantum phase transition can be studied as the function of
density $n$, and the  phase diagram is on 
Fig.~\ref{fig3}d.
The number of Dirac points in the ordered phase remains
the same.

{\em In conclusion}, we investigated the interplay of the trigonal
spectrum of the bare spectrum of the bilayer graphene with the
electron-electron interaction. The derived RG equations allowed us to
reveal the rich phase diagram \cite{elsewhere} determined by the few
(currently unknown) microscopic inputs. For a reasonably wide range
of the initial conditions \cite{footnote2} we found 
${\cal C}_{6v} \to {\cal C}_{2v}$
symmetry breaking and connected it with the change of the topology of
the single particle spectrum. We predicted the quantum phase
transition of the first order  as a function of
the electron density. Such a transition  should be most readily observed
in the hysteretic  dependence of the conductivity on the gate
controlled carrier density in the vicinity of the neutrality point.

This work was supported by US DOE contract No. DE-
AC02-06CH11357 (I.A) and by EPSRC EP/G041954/1  (V.F.).
We are grateful to A. Geim and A. Chubukov for valuable discussions.


\begin{thebibliography}{99}
\bibitem{Lifshitz} 
{
I. M. Lishitz, Zh. Exp. Teor. Fiz., 38, 1565 (1960) [Sov. Phys. JETP
11, 1130 (1960)]; A. A. Abrikosov, {\em Fundamentals of the Theory of
  Metals.}
 Elsevier, 1988.} 
\bibitem{Falko} E. McCann \& V. Fal'ko, Phys. Rev. Lett. {\bf 96}, 086805 (2006).
\bibitem{Yacoby} {B.E. Feldman, J. Martin, \& 
A. Yacoby, Nature Physics 5, 889 (2009).}

\bibitem{Geim} A. Geim (private communication).
\bibitem{Levitov1}R. Nandkishore \& L. Levitov, arXiv:0907.5395v1.
\bibitem{Levitov2}R. Nandkishore \& L. Levitov, arXiv:0907.5395v2,   
Phys. Rev. Lett., {\bf 104}, 156803, (2010).

\bibitem{Vafek}O.Vafek \& K.Yang,  Phys. Rev. B {\bf 81}, 041401(R) (2010). 

\bibitem{AKT} I.L. Aleiner, D.E.Kharzeev, \&
  A.M.Tsvelik, Phys. Rev. B {\bf 76}, 195415 (2007). 
\bibitem{Basko} D.M. Basko, Phys. Rev. B {\bf 78}, 125418 (2008).
\bibitem{Son}For similar treatment in monolayer see
  Ref.~\cite{AKT} and J.E. Drut \& D.T. Son, Phys. Rev. B {\bf 77}, 075115 (2008).

\bibitem{Footnote} Analytic expressions for coefficients $\alpha_{1,2,3}$ are
$
 \begin{pmatrix} \alpha_1;\alpha_2;\alpha_3\end{pmatrix}=
 \frac{1}{2\pi}\intop_{-\infty}^\infty
 \frac{dx D(x)}{(1+x^2)^2}\begin{pmatrix} \frac{1-3x^2}{(1+x^2)};2;
 \frac{D(x)}{4}
 \end{pmatrix},
 $
 where the function $D(x)$ is defined on Fig.~\ref{fig2}b.
\bibitem{elsewhere} Y. Lemonik, I.L. Aleiner, \& V.I. Fal'ko (in
  preparation).
\bibitem{footnote2} Line \req{glSol} is an unstable 
solution of \req{RGgs} and may be used only if the bare values
of all other $g_\cdot \lesssim 1/N$, so that they are still small
at energy ${\cal E}_1$. All possible stable directions, their
basins of attraction, and extremely rich phase diagram will be reported in Ref. \cite{elsewhere}.
\end{thebibliography}
\end{document}